\documentclass[preprint, amsmath,amssymb, superscriptaddress, prb]{revtex4-2}
\usepackage[dvipdfmx]{graphicx} 
\usepackage{dcolumn}
\usepackage{bm}
\usepackage{braket}
\usepackage{color}
\usepackage{ulem}
\usepackage{amsmath,amssymb}

\begin{document}
\title{Incommensurate magnetic order in an axion insulator candidate EuIn$_2$As$_2$ investigated by NMR measurement}

\author{Hikaru Takeda}
\email{takeda.hikaru@issp.u-tokyo.ac.jp}
\affiliation{Institute for solid state physics, University of Tokyo, 5-1-5 Kashiwanoha, Kashiwa, Chiba 277-8581, Japan}
\author{Jian Yan}
\affiliation{Institute for solid state physics, University of Tokyo, 5-1-5 Kashiwanoha, Kashiwa, Chiba 277-8581, Japan}

\author{Zhongzhu Jiang}
\affiliation{Key Laboratory of Materials Physics, Institute of Solid State Physics, HFIPS, Chinese Academy of Sciences, Hefei, 230031, China}
\affiliation{University of Science and Technology of China, Hefei, 230026, China}
\author{Xuan Luo}
\affiliation{Key Laboratory of Materials Physics, Institute of Solid State Physics, HFIPS, Chinese Academy of Sciences, Hefei, 230031, China}
\author{Yuping Sun} 
\affiliation{Key Laboratory of Materials Physics, Institute of Solid State Physics, HFIPS, Chinese Academy of Sciences, Hefei, 230031, China}
\affiliation{ Anhui Province Key Laboratory of Low-Energy Quantum Materials and Devices, HFIPS, High Magnetic Field Laboratory, Chinese Academy of Sciences, Hefei, 230031, China}
\affiliation{Collaborative Innovation Center of Advanced Microstructures, Nanjing University, Nanjing, 210093, China}

\author{Minoru Yamashita}
\affiliation{Institute for solid state physics, University of Tokyo, 5-1-5 Kashiwanoha, Kashiwa, Chiba 277-8581, Japan}

\date{\today}

\begin{abstract}
Magnetic topological insulators exhibit unique electronic states due to the interplay between the electronic topology and the spin structure. The antiferromagnetic metal $\rm{EuIn_2As_2}$ is a prominent candidate material in which exotic topological phases, including an axion insulating state, are theoretically predicted depending on the magnetic structure of the $\rm{Eu}^{2+}$ moments. 
Here, we report experimental results of the nuclear magnetic resonance (NMR) measurements of all the nuclei in $\rm{EuIn_2As_2}$ to investigate the coupling between the magnetic moments in the Eu ions and the conduction electrons in $\rm{In_2As_2}$ layers and the magnetic structure.
The $^{75} \rm{As}$ and $^{115}\rm{In}$ NMR spectra observed at zero external magnetic field reveal the appearance of internal fields of $4.9$ and $3.6\ \rm{T}$ respectively at the lowest temperature, suggesting a strong coupling between the conduction electrons in the $\rm{In_2As_2}$ layer and the ordered magnetic moments in the Eu ions. The $^{75}\rm{As}$ NMR spectra under in-plane external magnetic fields show broad distributions of the internal fields produced by an incommensurate fan-like spin structure which turns into a forced ferromagnetic state above $0.7\ \rm{T}$. We propose a spin reorientation process that an incommensurate helical state at zero external magnetic field quickly changes into a fan state by applying a slight magnetic field.
\end{abstract}
\maketitle

\vspace{8mm}
\noindent
{\bf Introduction}\label{introduction}

Nontrivial topological properties of electrons in a solid realize unconventional quantum transport, such as Dirac electrons on the surface of topological insulators (TIs) and the chiral edge current in the quantum Hall state of two-dimensional electrons under a strong external magnetic field \cite{Hasan2010}. Recently, various attempts have been made to find further quantum transports in a magnetic TI, in which conduction electrons with non-trivial band topologies are strongly coupled with a magnetic moment in neighboring layers \cite{Tokura2019, Bernevig2022, Wang2023b}. For example, the quantum anomalous Hall effect has been studied in thins films of chromium-doped $\rm{(Bi,Sb)_2Te_3}$ \cite{Chang2013, Xiao2018} and $\rm{MnBi_2Te_4}$ \cite{Deng2020, Liu2021b}. 

\begin{figure}
\includegraphics[width=10cm]{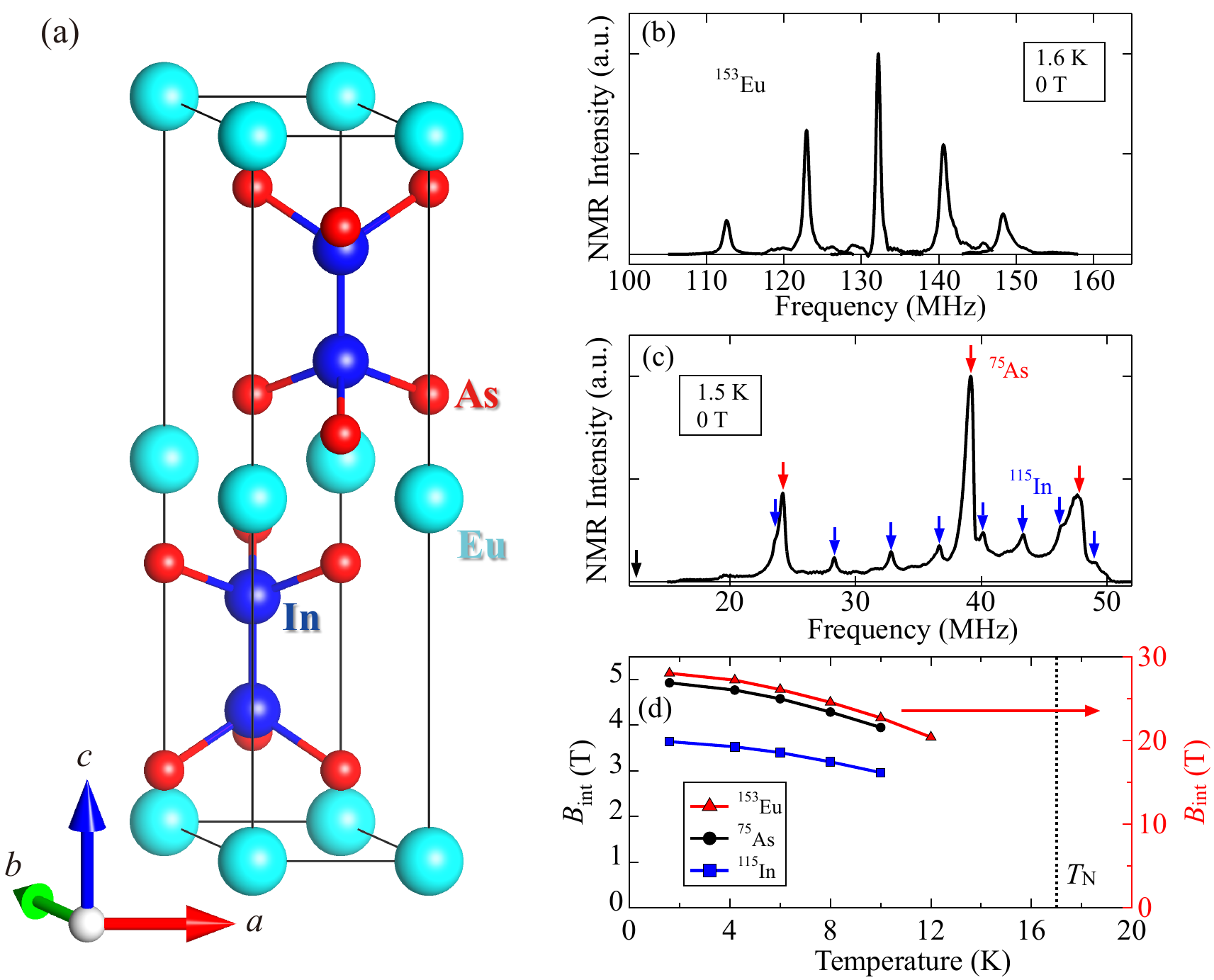}
\caption{\label{fig1} NMR signals of $\rm{EuIn_2As_2}$ at zero external magnetic field. (a) Crystal structure of $\rm{EuIn_2As_2}$ drawn by VESTA\cite{Momma2011}. The solid line represents the unit cell. (b) $^{153}\rm{Eu}$ NMR spectra observed at 1.6 K. (c) $^{75}\rm{As}$ and $^{115}\rm{In}$ NMR spectra observed at 1.5 K. The absorption peaks for each nucleus are pointed by red and blue arrows. The black arrow points to the frequency where another absorption peak for the $^{115}\rm{In}$ is expected to be detected. (d) Temperature dependence of the internal fields at $^{153}\rm{Eu}$ (right axis), $^{75}\rm{As}$ and $^{115}\rm{In}$ (left axis) nuclei. The solid lines are guides to eyes.}
\end{figure}

Recently, a series of Eu-based compounds have been intensively studied for exploring unusual topological states \cite{Bi2022,Du2022, Gui2019, Wang2019, Li2019b}.
The rare-earth Zintl compound $\rm{EuIn_2As_2}$ (Fig.\,\ref{fig1}(a)) \cite{Goforth2008b} is one promising candidate to realize these higher order topological states \cite{Xu2019}. In this compound, the spins of the Eu ions form an antiferromagnetic (AFM) order below the N\'{e}el temperature  $T_{\rm{N}} = 17\ {\rm{K}}$. In the AFM state, the ferromagnetically ordered magnetic moments parallel to the $ab$ plane in the single Eu layer \cite{Yan2022, Zhang2020a, Soh2023a, Riberolles2021, Gen2024, Donoway2023} align antiferromagnetically in the alternating stacking direction. The conduction electrons in the $\rm{In_2As_2}$ layers are sandwiched by the Eu layers, implying a strong coupling between the localized Eu moment and the conduction electrons. This coupling is pointed out to realize higher order topological states, including an axion insulating state in the collinear A-type AFM state, by the first-principles calculations \cite{Xu2019}. However, the strength of the coupling between the Eu moments and the conduction electrons has not been elucidated. 
In addition, although a double-$Q$ broken helix state has been suggested from the neutron diffraction \cite{Riberolles2021} and the resonant X-ray scattering (RXS) measurements \cite{Soh2023a, Gen2024}, the details of the broken-helix magnetic states remain controversial. For instance, the helical state with a commensurate $q$ vector is proposed by a RXS study \cite{Soh2023a}, while the state with an incommensurate $q$ vector is proposed by another RXS \cite{Gen2024} and the neutron diffraction \cite{Riberolles2021} studies. Furthermore, an unpinned broken helix state \cite{Donoway2023} or an incommensurate broken helix state with a sample dependence \cite{Gen2024} have been suggested by recent studies.

Here, we investigate the magnetic state of $\rm{EuIn_2As_2}$ by using NMR measurements. Our $^{153}\rm{Eu}$, $^{115}\rm{In}$, and $^{75}\rm{As}$ NMR measurements reveal that the ordered magnetic moments at the Eu ions yield the internal field of 28 T, 3.6 T, and 4.9 T at the $^{153}\rm{Eu}$, $^{115}\rm{In}$, and $^{75}\rm{As}$ nuclei respectively. The internal fields at the $^{115}\rm{In}$, and $^{75}\rm{As}$ nuclei are dominantly caused by transferred hyperfine interactions which are caused by a $c$-$f$ hybridization, suggesting  strong couplings between the magnetic moments at the Eu ions and the conduction electrons in the $\rm{In_2As_2}$ layers. We further find that an incommensurate magnetic state with a non-uniform semi-circle fan structure appears by applying an in-plane magnetic field. In addition, the reorientation process in the ordered moments under the external field, in which more spins are concentrated in the direction perpendicular to the applied field, enhances the NMR intensity. On the other hand, the magnetic structure smoothly transforms to the forced ferromagnetic state when the magnetic field is applied perpendicular to the plane. These results indicate an incommensurate magnetic structure weakly pinned in the crystal axes in this compound.

\vspace{8mm}
\noindent
{\bf Experimental Results}\label{results} \\
{\it NMR measurements at zero external magnetic field}\label{NMR_order}

In the magnetically ordered phase, spontaneous magnetic moments at the Eu sites generate large internal fields ${\bf{B}}_{\rm{int}}$ at the $^{153}\rm{Eu}$, $^{75}\rm{As}$ and $^{115}\rm{In}$ nuclei, which enables us to observe the NMR spectra even at zero external magnetic field. Figure\,\ref{fig1}(b) shows the  $^{153}\rm{Eu}$ NMR spectrum observed at 1.6 K.  Five absorption peaks are observed corresponding to the transitions between the nuclear spin energy levels of the $^{153}\rm{Eu}$ $(I=5/2)$. Note that the intensity of the spectrum is corrected by taking account the different spin echo decay time $T_2$ for each absorption peak. Figure\,\ref{fig1}(c) shows the $^{75}\rm{As}$ and $^{115}\rm{In}$ spectra observed at 1.5 K.  Three absorption peaks of $^{75}\rm{As}$ $(I=3/2)$ marked by red arrows and eight peaks of $^{115}\rm{In}$ $(I=9/2)$ by blue arrows are observed in the frequency range from $20\ \rm{MHz}$ to $50\ \rm{MHz}$. Another resonance line for the $^{115}\rm{In}$ is expected to exist at $12.6\ \rm{MHz}$. In contrast to $^{153}\rm{Eu}$, $T_2$ are almost the same for all the absorption peaks of $^{75}\rm{As}$ and $^{115}\rm{In}$ nuclei.  

We measured the temperature dependence of the spectra for $1.5\ {\rm{K}}\leq T \leq 12\ {\rm{K}}$. From the resonance frequencies of the absorption peaks, the temperature dependence of the absolute value of the internal field $B_{\rm{int}}$ for each nucleus are evaluated as shown in Fig.\,\ref{fig1}(d). In this evaluation, we use the  nuclear spin Hamiltonian expressed by Eq.\,(\ref{eq:hamiltonian_nuclear}) with adopting the recent experimental results that the ordered magnetic moments of Eu are in the $ab$ plane \cite{Soh2023a, Riberolles2021, Gen2024, Donoway2023}. The obtained absolute values of  the internal field $B_{\rm{int}}$ and the quadrupole frequency $\nu_Q$ (see Materials and Method) at the lowest temperature are listed in Table\,\ref{table1}.  The internal field at the $^{153}\rm{Eu}$ nucleus obtained in this analysis is in good agreement with the hyperfine field of 26.5 T evaluated by the $^{151}\rm{Eu}$ M\"{o}ssbauer spectroscopy at 5 K \cite{Riberolles2021}. The amplitude of this internal field at the $^{153}\rm{Eu}$ is also comparable to those for the Eu compounds composed of $\rm{Eu}^{2+}$ ion \cite{Ding2017, Higa2017, Ding2020}. 
Given that the dipole coupling ($\sim0.03\ {\rm{T/\mu_B}}$) is much smaller than the transferred hyperfine coupling ($\sim-0.73\ {\rm{T/\mu_B}}$) for the $^{75}\rm{As}$ nucleus as discussed in Supplementary Note 3B, the large internal fields of several Tesla at the $^{75}\rm{As}$ and $^{115}\rm{In}$ nuclei are dominantly caused by transferred hyperfine interactions between the ordered magnetic moments of Eu ions and the nuclei.

According to first-principles band structure calculations, the bands near the Fermi level are composed of As $4p$ and In $5s$ electrons at As and In ions, while fully spin-polarized $4f$ orbitals of Eu ion are pushed down below the Fermi level due to strong correlation effect \cite{Xu2019, Riberolles2021}. In this electronic structure, the localized $4f$ electrons couple with the conduction electrons via a $c$-$f$ hybridization, which results in the transferred hyperfine fields at the $^{75}\rm{As}$ and $^{115}\rm{In}$ nuclei. Therefore, the emergence of the internal fields of several Tesla suggests strong couplings between the $4f$ electrons and the conduction electrons through the hybridization. Although quantitative estimation of the coupling requires further theoretical studies, the internal fields obtained in our study should be  good measures for the strength of the coupling. It should be noted that this strong coupling between the $4f$ electrons and the conduction electrons is also shown by the sharp peak at $T_{\rm{N}}$ in the temperature dependence of the resistivity, the magneto resistance, and the anomalous Hall effects \cite{Yan2022}.

\begin{table}[b]
\caption{\label{table1} The absolute values of the internal fields $B_{\rm{int}}$ and the nuclear quadrupole frequencies $\nu_Q$ at $^{75}\rm{As}$, $^{115}\rm{In}$, and $^{153}\rm{Eu}$ nuclei at 1.5 and 1.6 K. }
\begin{ruledtabular}
\begin{tabular}{ccc}
Nucleus & $B_{\rm{int}}$ (T) & $\nu_Q$ (MHz) \\
\hline
$^{75}\rm{As}$ & 4.94 & 24.47 \\
$^{115}\rm{In}$ & 3.64 & 8.95 \\
$^{153}\rm{Eu}$ & 27.99 & 17.92 \\
\end{tabular}
\end{ruledtabular}
\end{table}

\vspace{8mm}
\noindent
{\it Magnetization and $^{75}\rm{As}$ NMR measurements in external magnetic fields}\label{MH}

\begin{figure}
\includegraphics[width=12cm]{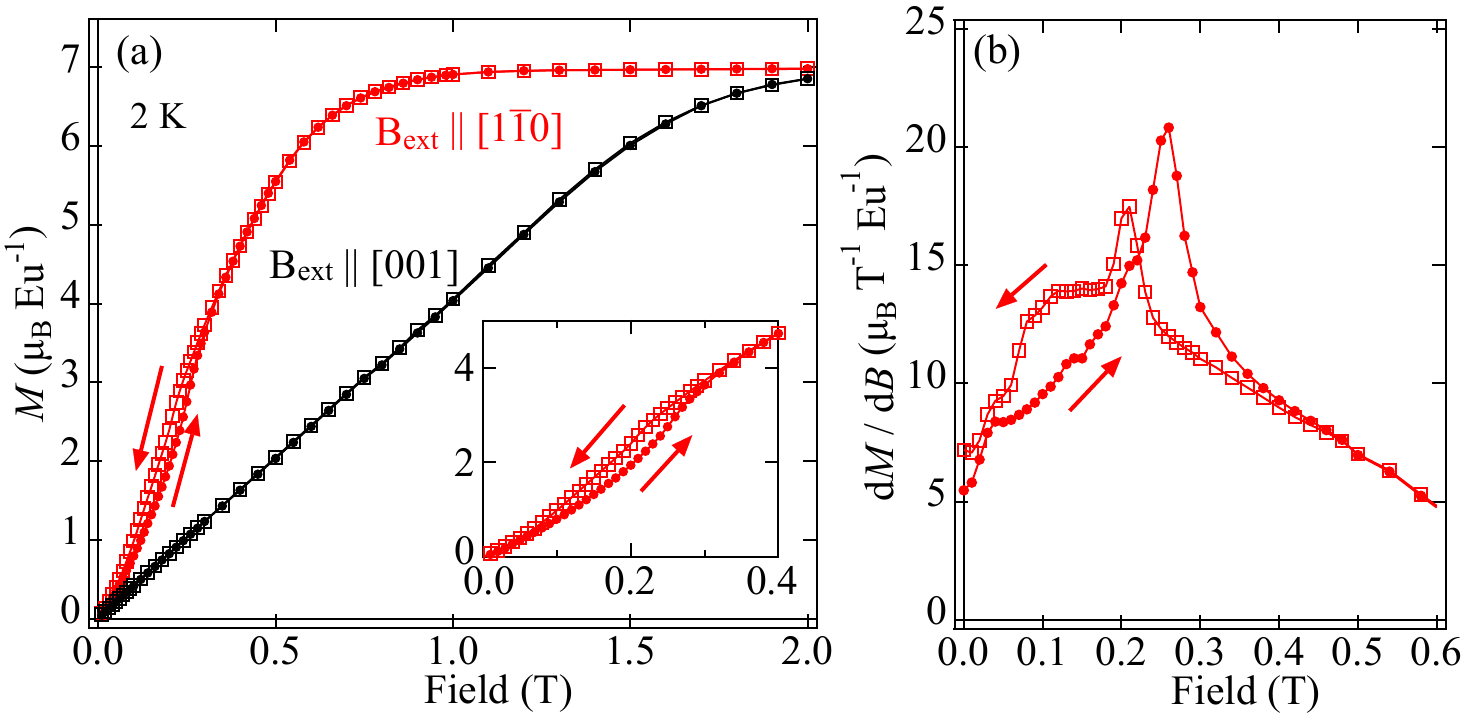}
\caption{\label{fig2} Magnetizations in the magnetically ordered state. (a) Magnetizations measured at 2 K in ${\bf{B}}_{\rm{ext}} \parallel [1\bar{1}0]$ and $[001]$. The solid circles (open squares) are the data measured while increasing (decreasing) magnetic field. The inset shows the enlarged magnetization curve for ${\bf{B}}_{\rm{ext}} \parallel [1\bar{1}0]$ around 0.2 T. (b) Field derivatives of the magnetizations for ${\bf{B}}_{\rm{ext}} \parallel [1\bar{1}0]$.}
\end{figure}

Figure\,\ref{fig2} shows the magnetization measured at 2 K in the external magnetic field applied ${\bf{B}}_{\rm{ext}}$ along $[1\bar{1}0]$ and $[001]$ directions. As previously reported \cite{Yan2022, Soh2023a, Riberolles2021, Gen2024, Zhang2020a}, the magnetization curves exhibit an easy-plane anisotropy as well as anomalies with a magnetic hysteresis in ${\bf{B}}_{\rm{ext}}\parallel [1\bar{1}0]$ for $0.05\ {\rm{T}}\lesssim B_{\rm{ext}}\lesssim 0.4\ {\rm{T}}$ which is more clearly seen in the external field derivative (Fig.\,\ref{fig2}(b)).

\begin{figure}
\includegraphics[width=12cm]{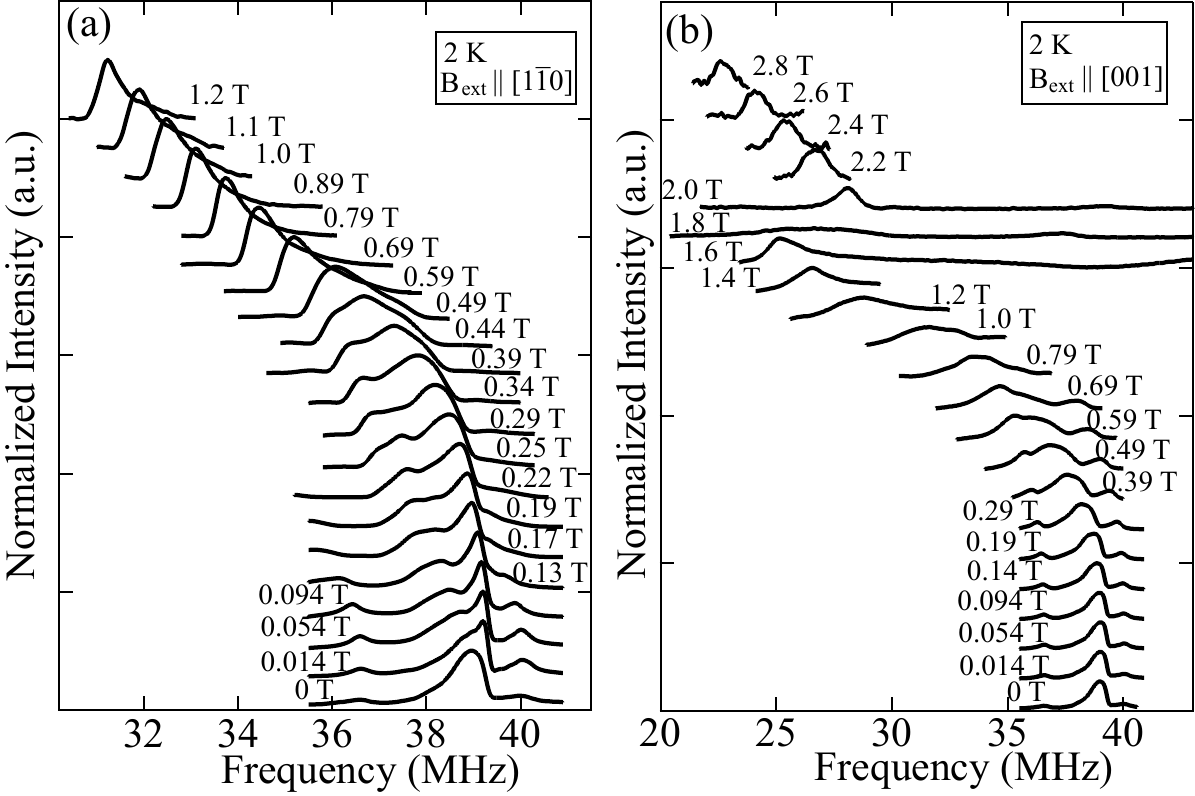}
\caption{\label{fig3} $^{75}\rm{As}$ NMR spectra in the magnetically ordered state. $^{75}\rm{As}$ NMR spectra observed at 2 K in (a) ${\bf{B}}_{\rm{ext}}\parallel [1\bar{1}0]$ and (b) ${\bf{B}}_{\rm{ext}}\parallel [001]$. The intensity of each spectrum is normalized for clarity.}
\end{figure}

In order to investigate the magnetic structure, we measured the center transition of the $^{75}\rm{As}$ NMR spectrum in external magnetic fields. Since the NMR spectrum is nothing but the histogram of the distributed local magnetic fields which are vector sums of the internal and external fields, we can reveal the distribution of the internal field from the spectral shape, which enables us to discuss the magnetic structure. Figure\,\ref{fig3}(a) and (b) show the spectra observed at 2 K in ${\bf{B}}_{\rm{ext}}\parallel [1\bar{1}0]$ and $[001]$ respectively, in which small absorption peaks from the $^{115}\rm{In}$ nucleus are also observed besides the $^{75}\rm{As}$ NMR spectrum. 

The application of an external field shifts the spectrum toward the lower frequency and changes the shape.   In ${\bf{B}}_{\rm{ext}}\parallel[1\bar{1}0]$, the spectrum has an asymmetric shape with a peak and a step for $0\ {\rm{T}} < B_{\rm{ext}} \lesssim 0.4\ {\rm{T}}$ and extends to the lower frequency with increasing the field. In the higher magnetic field, the peak and step are merged into a single peak, which results in an asymmetric spectrum with larger intensity at high frequencies above $\sim0.5\ {\rm{T}}$. 

The spectrum in ${\bf{B}}_{\rm{ext}} \parallel [001]$ simply broadens and shifts to lower frequency without notable changes in the spectral shapes up to 1.6 T (see Fig.\,\ref{fig3}(b)). This evolution of the spectrum can be explained by closing of an umbrella-type magnetic structure (see Supplementary Note 1).  The disappearance of the NMR spectrum at 1.8 T and the discontinuous change of the peak frequency above and below 1.8 T are caused by the change of the quantization axis of the nuclear spins, specific to the $^{75}\rm{As}$ NMR in this material. At the higher fields, the frequency shift of the observed spectrum is almost proportional to the external field, which suggests that all the ordered moments are forced to point to the direction of the external field.

\begin{figure}
\includegraphics[width=8cm]{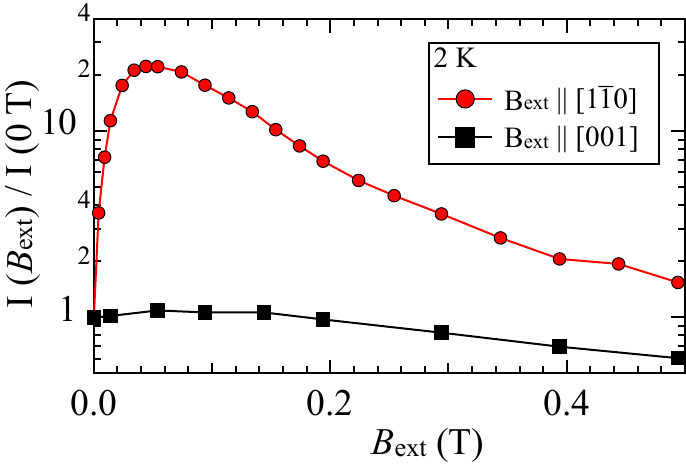}
\caption{\label{fig4} Magnetic-field dependence of the intensity of $^{75}\rm{As}$ NMR signal. Field dependence of the intensity of the $^{75}\rm{As}$ NMR spectra $I(B_{\rm{ext}})$ taken under  ${\bf{B}}_{\rm{ext}}\parallel [1\bar{1}0]$ (red) and (b) ${\bf{B}}_{\rm{ext}}\parallel [001]$ (black) normalized by that for zero field.}
\end{figure}

Note that the spectral intensities in Fig.\,\ref{fig3} are normalized for clarity because these intensities depend on the applied field as shown in Fig.\,\ref{fig4}. As shown in Fig.\,\ref{fig4}, while the intensity shows a slight change for ${\bf{B}}_{\rm{ext}}\parallel[001]$, that for ${\bf{B}}_{\rm{ext}}\parallel[1\bar{1}0]$ shows drastic enhancement up to 0.05 T, which is followed by a gradual decrease at higher fields.

\vspace{8mm}
\noindent
{\bf Discussion}\label{discussion}

\begin{figure}
\includegraphics[width=8cm]{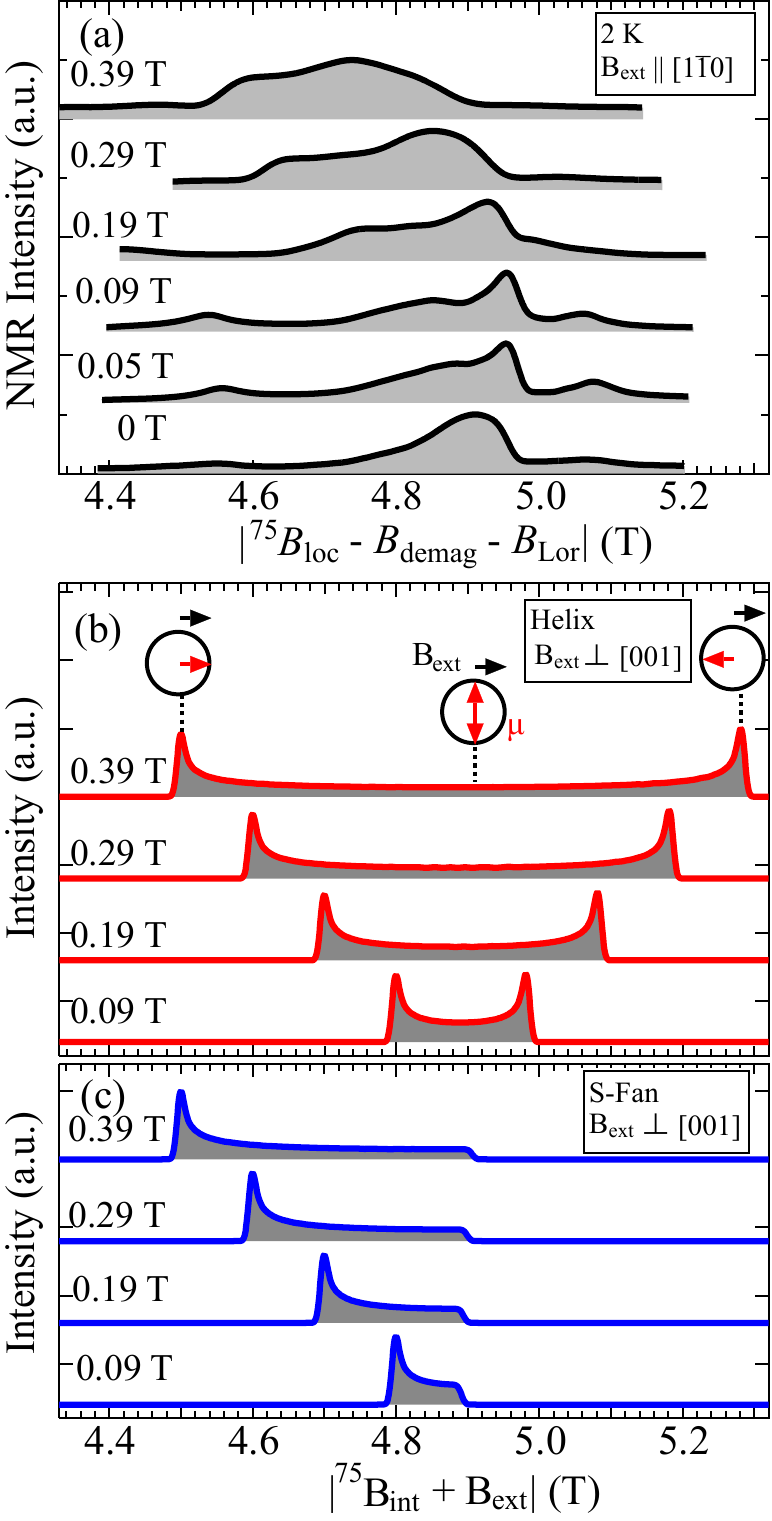}
\caption{\label{fig5} Comparison between the observed and the calculated $^{75}\rm{As}$ NMR spectra. (a) The external magnetic field evolution of the observed $^{75}\rm{As}$ NMR spectra in ${\bf{B}}_{\rm{ext}}\parallel [1\bar{1}0]$. The spectra are plotted against the absolute value of the local magnetic field subtracted by demagnetization and Lorentz cavity fields. [(b), (c)] The histograms of the distributed local fields for (b) an incommensurate helix state and (c) an incommensurate semi-circle fan state with the uniform spin distribution in some in-plane external magnetic fields. The inset in (b) represents the schematic figures for the orientation of the ordered moments which produce the local fields pointed by the dotted lines.}
\end{figure}

As shown in Fig.\,\ref{fig3}(a), in-plane external magnetic fields drastically change the shape of the $^{75}\rm{As}$ NMR spectrum, suggesting that the local field distribution is altered. To discuss the field evolution of the local field, we focus on the NMR spectra at representative external fields as shown in Fig.\,\ref{fig5}(a). Note that the spectra in Fig.\,\ref{fig5}(a) are plotted as a function of $|^{75}B_{\rm{loc}}-B_{\rm{demag}}-B_{\rm{Lor}}|$ ($=|{\bf{B}}_{\rm{ext}}+^{75}{\bf{B}}_{\rm{int}}|$), in which $^{75}B_{\rm{loc}}$, $B_{\rm{demag}}$, $B_{\rm{Lor}}$, and $^{75}{\bf{B}}_{\rm{int}}$ are the local magnetic field at the $^{75}\rm{As}$ nucleus, demagnetization and Lorentz cavity fields and the internal magnetic field, respectively (see Materials and Method). The local magnetic field  $^{75}B_{\rm{loc}}$ is estimated from the resonance frequency by using Eq.\,(\ref{eq:hamiltonian_nuclear}) and the experimental results that the local fields lie in the $ab$ plane \cite{Soh2023a, Riberolles2021, Gen2024, Donoway2023}.

The observed spectra shown in Fig.\,\ref{fig5}(a) have the following two important features. First, the spectra observed in the external magnetic fields show broad distributions of the intensity with a high-field peak and a low-field step, rather than consist of discrete peaks as observed in a commensurate magnetic state (see Supplementary Note 2). From this feature, we can exclude a commensurate magnetic structure as the ordered state under the applied field. It is also quite unlikely that a commensurate magnetic structure, such as the broken helix suggested in Ref.\,\onlinecite{Soh2023a}, is transformed to an incommensurate one under the weak in-plane field. Therefore, it is more likely that an incommensurate magnetic structure appears at zero-field in this compound.

\begin{figure}
\includegraphics[width=10cm]{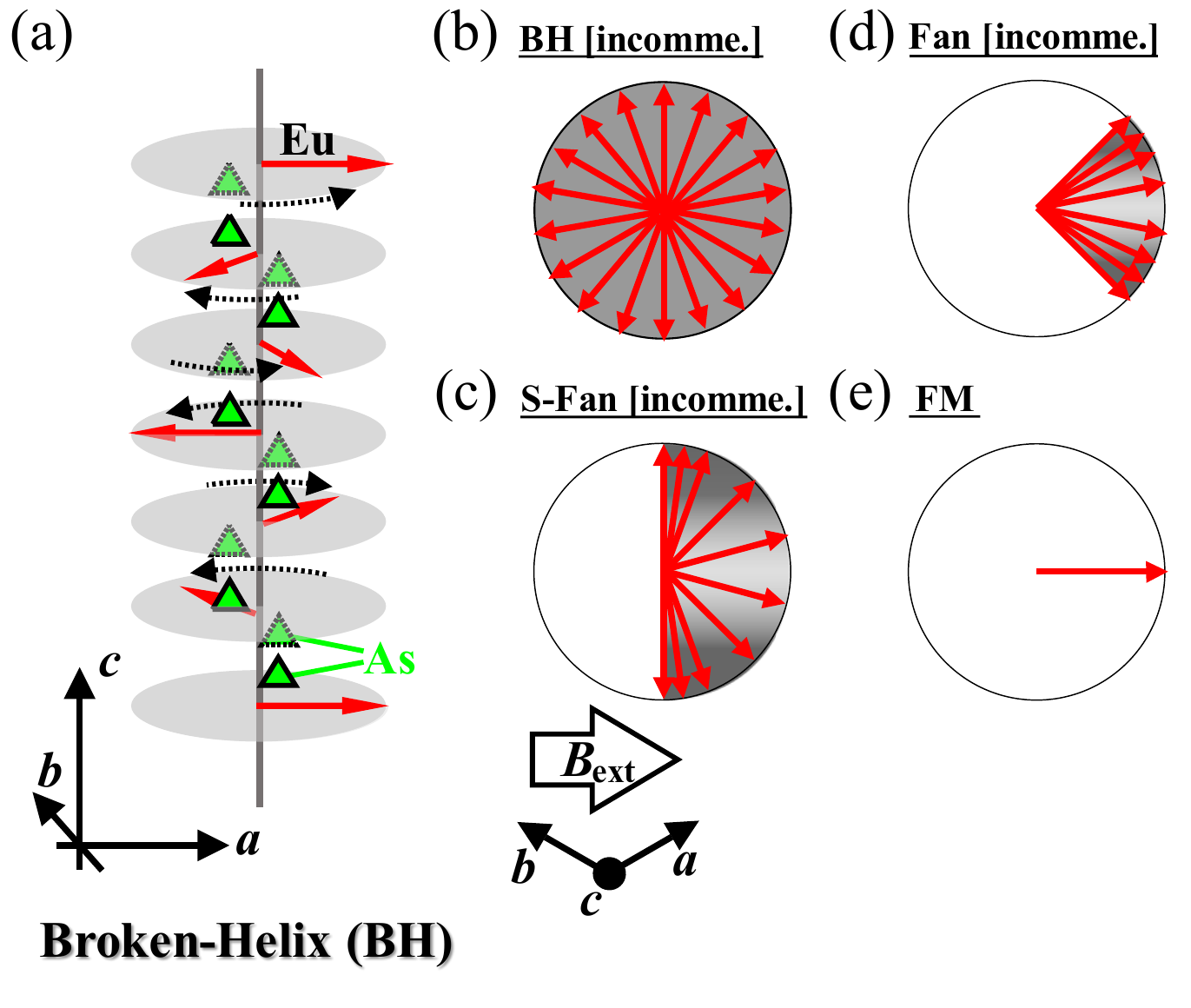}
\caption{\label{fig6} Magnetic-field dependence of the magnetic structure. (a) Schematic figure of the magnetic structure of the broken helix state. The red arrows on the gray circles denote the magnetic moments lying on the helical plane at the Eu sites, while the green triangles are As sites beside the Eu layers. The dotted arrows represent the rotation of the magnetic moment from the direction of the magnetic moment in the lower layer. (b)-(e) Schematic illustrations of incommensurate magnetic structures viewed along the $c$ axis, which are evolved by the in-plane magnetic field (${\bf{B}}_{\rm{ext}}\parallel [1\bar{1}0]$) from a broken helix (BH, b) to a semi-circle fan (S-Fan, c), a fan (Fan, d), and a forced ferromagnetic (FM, e) state. The density of the red arrows and gray gradations represent the distribution of the ordered magnetic moments in the broken helix, semi-circle fan and a fan state.}
\end{figure}

Second, the spectrum extends to the lower field side with increasing the external field, while the high-field peak shifts modestly toward lower field, which indicates that the internal field is not distributed in the same direction to the external field. To see this, let us show the calculated histogram for the incommensurate broken helix structure shown in Figs.\,\ref{fig6}(a) and (b). In this state the ordered moments at the Eu ions in an Eu layer can be given by
\begin{equation}{\label{eq:magnetization}}
{\bf{\mu}}=\mu(\cos \theta, \sin \theta, 0),
\end{equation}
where $\theta$ denotes the angle between the ordered moment and the $a$ axis. The $^{75}\rm{As}$ nuclei beside this Eu layer experience the internal field which is approximately given by
\begin{equation}{\label{eq:Bint_heli}}
^{75}{\bf{B}}_{\rm{int}}=(^{75}{\bf{A}}_{\rm{NN}}+{\bf{\alpha}}^{\rm{NN}}_{\rm{dip}})\cdot{\bf{\mu}},
\end{equation}
where $^{75}{\bf{A}}_{\rm{NN}}$ and ${\bf{\alpha}}^{\rm{NN}}_{\rm{dip}}$ are the transferred hyperfine and dipole coupling tensors connecting the ordered moment and the internal field, respectively. As shown in Supplementary Notes 3, 4, and 5 where the evaluation of the coupling tensors are demonstrated, these two coupling tensors are uni-axial, which have the in-plane components $^{75}A_{\rm{NN}\perp}=-0.73\ \rm{T/\mu_B}$ and $\alpha^{\rm{NN}}_{\rm{dip}\perp}=0.03\ \rm{T/\mu_B}$. Since the angle $\theta$ differs from each other in each Eu layer and is distributed uniformly from 0 to $2\pi$, the orientation of the internal field at the $^{75}\rm{As}$ nuclei is also distributed uniformly in the $ab$ plane. Then, the local field given by the vector sum of the internal field and the external one, $^{75}{\bf{B}}_{\rm{loc}}=^{75}{\bf{B}}_{\rm{int}}+{\bf{B}}_{\rm{ext}}$ yields the double-horn type histogram of the local field as shown in Fig.\,\ref{fig5}(b). The two peaks are attributed to the $^{75}\rm{As}$ nuclei where the internal field follows the relation $\pm{^{75}\bf{B}}_{\rm{int}}\parallel{\bf{B}}_{\rm{ext}}$. The interval $\Delta B$ of the two peaks increase as ${\bf{B}}_{\rm{ext}}$ increases, following $\Delta B\sim 2B_{\rm{ext}}$. In contrast to the histogram in Fig.\,\ref{fig5}(b), the high-field component is absent in the observed spectrum as shown in Fig.\,\ref{fig5}(a). Because the coupling constant is negative, the absence in the high-field component in the observed spectrum indicates that the ordered moment pointing to the opposite direction to the applied field quickly disappears by applying the external field. Thus, one needs to consider a change of the magnetic structure to a fan-like state under an applied field.

Figure\,\ref{fig5}(c) shows a calculated histogram of the local field for a semi-circle fan state with the uniform spin distribution. As shown in Fig.\,\ref{fig5}(c), the uniform distribution of the spins in the semi-circle fan gives the spectrum with the low-field peak and the high-field step. The low-field peak shifts to lower field at a higher external field, whereas the high-field step does not shift. Although the observed spectrum extends to the lower field side by applying an external field as the calculated histogram of the semi-circle fan does, the relative intensity of the high- and the low-field edges in the calculated histogram is opposite to the observed spectrum. Therefore, the ordered moments are inferred to have a non-uniform distribution with more spins pointing perpendicular to the applied field as illustrated by the grey gradation in Fig.\,\ref{fig6}(c).

This non-uniform distribution of the spins in the semi-circle fan structure is also consistent with the field enhancement of the NMR intensity (Fig.\,\ref{fig4}). Normally, an NMR intensity, which reflects the population difference between the neighboring energy levels, is hardly increased by the applied magnetic field because the energy splitting of the nuclear spins under an applied field is negligibly smaller than the thermal energy. However, the NMR intensity could be enhanced via the hyperfine interaction. 

Let us first consider the NMR for the ferromagnet with a weak spin anisotropy to see the general mechanism for the enhancement of the NMR signal. In the ferromagnet, the nuclear spins are excited by not only the external RF field $H_1$ but also the hyperfine field following the electronic moment response to $H_1$, which results in the enhancement of $H_1$. The enhancement factor is given by the ratio of the hyperfine field produced by the electronic moment to the restoring one which characterizes the torque experienced by the electronic moment when it is tilted from its position at rest by $H_1$ \cite{Meny2021a}. With respect to the received NMR signal, it is also enhanced by the same factors, because the electronic magnetization, which is driven into rotation by the motion of the nuclear magnetization via the hyperfine interaction, contributes to the signal. This phenomenon is known as a domain rotation mechanism for the NMR enhancement in ferromagnets \cite{Meny2021a}.

Although $\rm{EuIn_2As_2}$ is an antiferromagnet, the $^{75}\rm{As}$ NMR signal is enhanced at zero external magnetic field due to the motion of the net electronic magnetization induced by the RF field. 
In an A-type antiferromagnet such as $\rm{EuIn_2As_2}$, this enhancement can be further increased by applying an in-plane external magnetic field, which is explained by the following scenario reported in another A-type antifferomagnet $\rm{CrCl_3}$ \cite{Narath1963}. When a weak in-plane external field is applied in the A-type antiferromagnet with a small in-plane anisotropy, the net electronic magnetization is tilted from the applied field direction due to the weak in-plane magnetic anisotropy, which results in damping of the rotation of the net magnetization by the RF field. As the applied field increases, the net magnetization increases until the net magnetization turns parallel to the applied field by overcoming the anisotropy. In this situation, the enhancement factor increases as the magnetization increases, because the hyperfine field generated by the magnetization increases while the AFM coupling between the magnetic layers (i.e. the Eu layers for $\rm{EuIn_2As_2}$) acts as a restoring field which is independent of the external field. Beyond the critical field at which the net magnetization turns parallel to the applied field, the enhancement decreases because the restoring field is replaced by the external field. Note that the NMR signal enhancement by the application of the external magnetic field is not observed in ${\bf{B}}_{\rm{ext}}\parallel [001]$, because the spins in the $ab$ plane do not exhibit the redistribution which enhances the net magnetization. 

Thus, the enhancement of the NMR intensity up to 0.05 T can be explained by the presence of a small in-plane magnetic anisotropy in the $ab$ plane of $\rm{EuIn_2As_2}$ and the redistribution of the Eu moment perpendicular to the applied field. This field evolution of the ordered moments is compatible with the field dependence of the NMR spectrum. As shown in Fig.\,\ref{fig3}(a), the NMR spectrum first shows a quick increase of the high-frequency peak up to 0.05 T. Above 0.05 T, the high frequency peak gradually shifts to the lower frequency, which suggests that the fan-like spin distribution closes as shown in Fig.\,\ref{fig6}(d). In addition, the kink feature at 0.05 T in the magnetization curve (Fig.\,\ref{fig2}(b)) is compatible with this spin-flop-like redistribution. 

The weak magnetic anisotropy in the $ab$ plane demonstrated above suggests that the ordered moments are hardly pinned in the crystal axes, which excludes the presence of the magnetic domains \cite{Soh2023a} in an applied magnetic field. Recent experimental studies with RXS \cite{Gen2024} and optical probe \cite{Donoway2023} have proposed such an unpinned magnetic structure as well as the quick reorientation of the ordered moments by the slight in-plane field \cite{Gen2024}, which are consistent with our experimental results. Combined with the incommensurate periodicity found by our NMR measurements, it is inferred that an incommensurate broken helix state with weakly pinned ordered moments is realized in the magnetically ordered state as suggested in the previous studies \cite{Donoway2023, Gen2024}, which enables us to discuss the possibility of the emergence of the topological phase in $\rm{EuIn_2As_2}$. In $\rm{EuIn_2As_2}$ the axion insulator phase is expected to be realized in the presence of $\tau C_2$ symmetry, where $\tau$ and $C_2$ are the time reversal and two fold-rotational operators, in the magnetic structure \cite{Riberolles2021, Soh2023a}. The commensurate broken helix state with a six-layered period could satisfy this symmetry, when the ordered moments are pinned in the crystal lattice \cite{Riberolles2021, Soh2023a}. 
However, the broken helix state clarified in our study does not satisfy these two conditions, indicating that fine controls on the magnetic state are necessary to realize the axion insulator state in this compound.
Furthermore, tuning the Fermi level is required to realize the insulating phase, since the prestine sample of $\rm{EuIn_2As_2}$ is reported to be a hole-doped metal. A suppression of the hole band by tuning the Fermi level \cite{Yan2024} might change the magnetic interaction between Eu moments through the $c$-$f$ hybridization, which could give rise to a change of the helical magnetic structure of this material. Such a change in the electronic state would be detected as changes in the transferred hyperfine field at the $\rm{^{75}As}$ and $\rm{^{115}In}$ nuclei. Finally, we note that only the static properties of the magnetic state have been investigated in this study. For further understanding of the magnetic state, investigation of the spin dynamics by nuclear spin relaxation measurements is required, which remains as an important future issue.

In summary, the internal fields produced by the ordered magnetic moments at the Eu ions are detected via $^{153}\rm{Eu}$, $^{115}\rm{In}$ and $^{75}\rm{As}$ NMR spectra. The $^{75} \rm{As}$ and $^{115}\rm{In}$ nuclei experience the fields of $4.9$ and $3.6\ \rm{T}$ respectively at the lowest temperature, suggesting a strong coupling between the conduction electrons in the $\rm{In_2As_2}$ layer and the ordered moments. The $^{75}\rm{As}$ NMR spectra under in-plane external magnetic fields show broad distributions of the internal fields produced by the incommensurate fan-like spin structure which changes into a forced ferromagnetic state above $0.7\ \rm{T}$. This fan state is evolved from an incommensurate helical state by applying a slight in-plane magnetic field, which suggests the ordered moments are weakly pinned in the crystal lattice. Our results reveal the magnetic structure from a local perspective in a complementary way to the scattering measurements, providing crucial information to discuss the emergence of the topological phase which requires a specific symmetry in the magnetic structure.

\vspace{8mm}
\noindent
{\bf Methods}\label{procedure}

The single crystals of $\rm{EuIn_2As_2}$ were prepared by flux method. In our experiments, we used two pieces of the samples, $\#1$ and $\#2$, the sizes of which are $\sim0.5 \times 1.2\times 0.1 {\rm{mm}}^3$ and $\sim1.0 \times 1.5 \times 0.1 {\rm{mm}}^3$ respectively. For the sample\,$\#1$, we measured the magnetization at 2 K by using a SQUID magnetometer (Quantum Design MPMS), and $^{75}\rm{As}$, $^{115}\rm{In}$, and $^{153}\rm{Eu}$ NMR spectra in the magnetically ordered phase, while we used the sample\,$\#2$ to measure the $^{75}\rm{As}$ NMR spectra in the paramagnetic phase (see Supplementary Note 3). The frequency swept NMR spectra were obtained by summing the Fourier transform of spin echo signal recorded at equally spaced frequencies. Since all the nuclei in $\rm{EuIn_2As_2}$ have single crystallographical site in the crystal structure, the NMR spectrum of each nucleus shows absorption peaks determined by the nuclear spin $I = 3/2$, $9/2$ and $5/2$ for $^{75}\rm{As}$, $^{115}\rm{In}$, and $^{153}\rm{Eu}$, respectively as shown in Fig.\,\ref{fig1}(b)-(d).

In general, the NMR frequency is determined by the nuclear spin Hamiltonian,
\begin{equation}{\label{eq:hamiltonian_nuclear}}
H_I=h\gamma {\bf{I}}\cdot {\bf{B}}_{\rm{loc}}+H_Q,
\end{equation}
where the first term is the Zeeman interaction expressed with the Planck's constant $h$, the nuclear gyromagnetic ratio $\gamma$ $(^{75}\gamma=7.290\ {\rm{[MHz/T]}}, ^{115}\gamma=9.330\ {\rm{[MHz/T]}}, ^{153}\gamma=4.675\ {\rm{[MHz/T]}})$, and the nuclear spin $\bf{I}$, and the second term is electric quadrupole interaction. 
As shown in Supplementary Note 2, ${\bf{B}}_{\rm{loc}}$ is the local field at the nucleus, which is composed of the external ${\bf{B}}_{\rm{ext}}$, demagnetization ${\bf{B}}_{\rm{demag}}$, Lorentz cavity ${\bf{B}}_{\rm{Lor}}$, hyperfine ${\bf{B}}_{\rm{hf}}$ and classical dipole ${\bf{B}}_{\rm{dip}}$ fields,
\begin{equation}{\label{eq:local_field}}
{\bf{B}}_{\rm{loc}}={\bf{B}}_{\rm{ext}}+{\bf{B}}_{\rm{demag}}+{\bf{B}}_{\rm{Lor}}+{\bf{B}}_{\rm{hf}}+{\bf{B}}_{\rm{dip}}. 
\end{equation}

The quadrupole interaction is expressed as
\begin{equation}{\label{eq:efg}}
H_Q=\sum {\bf{V}}_{\alpha \beta}{\bf{Q}}_{\alpha \beta},
\end{equation}
where ${\bf{V}}_{\alpha \beta}=(\partial ^2 V)/(\partial x_{\alpha}\partial x_{\beta})$ is the electric field gradient (EFG) and ${\bf{Q}}_{\alpha \beta}=\frac{eQ}{6I(2I-1)} \{3/2 (I_{\alpha} I_{\beta}+I_{\beta} I_{\alpha} )-\delta_{\alpha \beta}I(I+1)\}$ is the nucleus quadrupole moment. In the case of $\rm{EuIn_2As_2}$, $H_Q$ for all the nuclei can be rewritten with the quadrupole frequency $\nu_Q=\frac{3eQ}{h2I(2I-1)}V_{zz}$ as,
\begin{equation}{\label{eq:efg2}}
H_Q=\frac{h\nu_Q}{6\{3I_z^2-I(I+1)\}},
\end{equation}
since all the nuclei have axial point symmetries \cite{Goforth2008b}.

\vspace{8mm}
\noindent
{\bf Author contributions}

H.T., J.Y., and M.Y. conceived the project. H.T. and J.Y. performed the NMR measurements on the sample grown and characterized by J.Y., Z.J., X.L., and Y.S. 
H.T. and M.Y. wrote the manuscript based on the input from all coauthors. H.T. and J.Y. are equally contributed to the work.

\vspace{8mm}
\noindent
{\bf Competing Interests}

The authors declare no competing interests. 

\vspace{8mm}
\noindent
{\bf Data availability}

The data that support the findings of this study are available upon reasonable request.

\vspace{8mm}
\noindent
{\bf Acknowledgments}

This work was supported by JSPS KAKENHI Grant No. JP22KF0111 (M.Y.) and No. JP23H01116 (M.Y.), National Key R$\&$D Program Grant No. 2023YFA1607402 (Z.J., X.L. and Y.S.) and No. 2021YFA1600201 (Z.J., X.L. and Y.S.),  National Natural Science Foundation of China Grant No. U2032215 (Z.J., X.L. and Y.S.), No. U1932217 (Z.J., X.L. and Y.S.), and No. 12274412 (Z.J., X.L. and Y.S.), and Systematic Fundamental Research Program Leveraging Major Scientific and Technological Infrastructure, Chinese Academy of Sciences under contract No. JZHKYPT-2021-08 (Z.J., X.L. and Y.S.). The authors appreciate M. Gen for fruitful discussions.

\newpage
\bibliographystyle{naturemag}

\newpage

\end{document}